# New exact solutions for the evaporation flux density of a small droplet on a flat horizontal substrate with a contact angle in the range of 135-180º

Peter Lebedev-Stepanov

FSRC "Crystallography and Photonics" RAS, Leninskii pr-t, 59, Moscow, 119333, Russia

National Research Nuclear University MEPhI, Kashirskoye shosse, 31, Moscow, 115409, Russia

Electronic mail: petrls@yandex.ru

Previously [Lebedev-Stepanov P.V. ArXiv:2103.15582v3], an expression was proposed for the evaporation flux density of a small liquid droplet having the shape of an axisymmetric spherical segment deposited on a horizontal substrate. The dependence of the flux density on the polar angle was established for arbitrary contact angles. This formula has the form of an integral and is rather complicated for use in modeling algorithms. An approximate expression was obtained for the evaporation flux density at small contact angles. However, the question of which simplified formulas should be appropriate to apply in other ranges of contact angles, for example, in the case of obtuse angles remains open. In this paper, we propose new exact solutions for the set of discrete contact angles $\theta = \pi\left(1 - \frac{1}{2j}\right)$, where $j=1,2,3…$  As an example, very simple exact expressions are obtained explicitly for the evaporation flux density for droplets with contact angles $\frac{3\pi}{4} = 135º$ and $\frac{5\pi}{6} = 150º$ that do not contain integral dependencies. They can also be used as approximate solutions for a narrow range of contact angles around the specified values.

## 1. Introduction

Previously, exact analytical solutions have been obtained to describe the evaporation rate and evaporation flux density of a small sessile liquid droplet having the shape of an axisymmetric spherical segment deposited on a horizontal substrate (Fig. 1) [1–4].

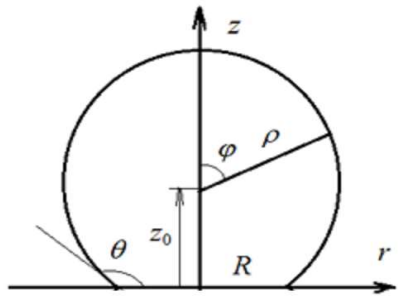

**Figure 1.** The geometry of the sessile droplet: $\theta$ is a contact angle, $\varphi$ is a polar angle, $\rho$ is a spherical segment radius, $R$ is the radius of contact line, $(z, r)$ are the cylindrical coordinates.

For the total evaporation rate, the following expression was obtained in [2]:

$$W = \pi R D(n_s - n_\infty) \left[ \frac{\sin\theta}{1+\cos\theta} + 4\int_0^\infty \frac{1+\cosh 2\theta\tau}{\sinh 2\pi\tau} \tanh[(\pi-\theta)\tau] d\tau \right]. \qquad (1)$$

where $D$ is a diffusion coefficient of the vapor in the air, $n$ is a vapor volume concentration outside the drop with the boundary conditions $n = n_S$ at the drop air-liquid surface and $n = n_\infty$ far from the drop, $R$ is the radius of contact line.

Two alternative expressions are obtained for the evaporation flux density. The first expression [2,3] was found by

$$J(\alpha) = D \frac{n_s - n_\infty}{R} \times$$
$$\times \left[ \frac{\sin\theta}{2} + \sqrt{2}(\cosh\alpha + \cos\theta)^{3/2} \int_0^\infty \frac{\cosh\theta\tau}{\cosh\pi\tau} \tanh[(\pi-\theta)\tau] P_{-1/2+i\tau}(\cosh\alpha)\tau d\tau \right], \qquad (2)$$

where $P_{-1/2+i\tau}(\cosh\alpha) = \frac{2}{\pi}\cosh\pi\tau \int_0^\infty \frac{\cos\tau t}{\sqrt{2(\cosh t + \cosh\alpha)}} dt$ is the Legendre function of the first kind. Here, toroidal coordinate $\alpha$ ranges in the interval from 0 (top of the drop) to $\infty$ (contact line). So, this coordinate is related to the cylindrical coordinate $r$ by

$$r = \frac{R\sinh\alpha}{\cosh\alpha + \cos\theta}, \qquad (3)$$

where $\theta$ is the contact angle of the droplet (Fig.1). Formula (2) is a double integral, being an implicit function of cylindrical coordinate $r$, which makes formula (2) extremely difficult to use in calculations. The second expression was proposed in [1,4]. It is an equivalent of (2) but has a simpler from a computational point of view. It allows to calculate the flux density as a function of the polar angle $\varphi$ explicitly (Fig. 1):

$$J(\varphi) = \frac{D(n_S - n_\infty)}{R} f(\varphi), \qquad (4)$$

where

$$f(\varphi) = \frac{\pi}{(\pi-\theta)^2} \frac{\sin^3\theta}{\cos\varphi - \cos\theta} \int_\varphi^\theta \frac{(1-\cos(\theta+\beta))^{\frac{\pi}{2(\pi-\theta)}} - (1-\cos(\theta-\beta))^{\frac{\pi}{2(\pi-\theta)}}}{\left((1-\cos(\theta+\beta))^{\frac{\pi}{2(\pi-\theta)}} + (1-\cos(\theta-\beta))^{\frac{\pi}{2(\pi-\theta)}}\right)^2} \frac{(\cos\beta - \cos\theta)^{\frac{\pi}{2(\pi-\theta)} - \frac{1}{2}}}{\sqrt{\cos\varphi - \cos\beta}} d\beta. \qquad (5)$$

From the point of view of application in computer simulations, this universal analytical expression that describes the evaporation flux density over the entire range of contact angles 0-180°, is still quite complex. It requires significant computational resources. To accelerate the calculations of the evaporation flux density, it is reasonable to use simplified approximate expressions.

For example, there is a very good approximation for the integral evaporation flux proposed by Picknett and Bexon [5]:

$$W = 2\pi\rho D(n_s - n_\infty) g(\theta), \qquad (6)$$

where

$$g(\theta) = 0.6366\theta + 0.09591\theta^2 - 0.06144\theta^3, \ 0 \leq \theta \leq 0.175;$$
$$g(\theta) = 0.00008957 + 0.6333\theta + 0.1160\theta^2 - 0.08878\theta^3 + 0.01033\theta^4, \ 0.175 \leq \theta \leq \pi. \quad (7)$$

This expression has a maximum error of about 0.2% and looks much more preferable for simulation than the exact analytical solution (1). Similarly, for the evaporation flux density, instead of the exact formulas (2) or (4), approximate expressions can be proposed for selected narrow ranges of droplet contact angles. So, earlier an expression for the evaporation flux density was proposed, applicable in the case of small contact angles [6]:

$$J(r) = J_0(\theta)\left(1 - \frac{r^2}{R^2}\right)^{-\lambda(\theta)}, \ \lambda(\theta) = \frac{1}{2} - \frac{\theta}{\pi}, \quad (8)$$

$$J_0(\theta) = \frac{D(n_s - n_\infty)}{R}(0.27\theta^2 + 1.30)(0.6381 - 0.2239(\theta - 0.25\pi)^2).$$

This expression gives a good description for contact angles smaller than 30°.

However, the question of which simplified formulas would be appropriate to apply in other ranges of contact angles, for example, typical for drops on hydrophobic substrates, still remains open. In this paper a solution to this problem is proposed.

## 2. Exact solutions for some values of contact angles

Previously [1,4], an expression was obtained that describes the vapor concentration near an evaporating drop:

$$n(\omega,\xi) = n_s - \frac{2(n_s - n_\infty)}{\pi - \theta}(\cosh\omega - \cos\xi)^{1/2} \cos\frac{\pi\xi}{2(\pi-\theta)} \int_{\varsigma=\omega}^{\infty}\left(\cosh\frac{\pi\varsigma}{(\pi-\theta)} - \cos\frac{\pi\xi}{(\pi-\theta)}\right)^{-1} \frac{\sinh\frac{\pi\varsigma}{2(\pi-\theta)}d\varsigma}{\sqrt{\cosh\varsigma - \cosh\omega}}. \quad (9)$$

This expression can also be represented as:

$$n(\omega,\xi) = n_s - \frac{(n_s - n_\infty)}{2\gamma}(\cosh\omega - \cos\xi)^{1/2}\int_\omega^\infty\left[\frac{\sinh\frac{\pi\varsigma}{2\gamma}}{\cosh\frac{\pi\varsigma}{2\gamma} - \cos\frac{\pi\xi}{2\gamma}} - \frac{\sinh\frac{\pi\varsigma}{2\gamma}}{\cosh\frac{\pi\varsigma}{2\gamma} + \cos\frac{\pi\xi}{2\gamma}}\right]\frac{d\varsigma}{\sqrt{\cosh\varsigma - \cosh\omega}}. \quad (10)$$

Here $\gamma = \pi - \theta$.

The integral (10) can be represented as the sum of a finite number of terms for some specific contact angles [7]. It was shown that expression (10) under the condition

$$\xi = \frac{\pi}{2j}, \text{ where } j=1,2,3,\ldots \quad (11)$$

can be rewritten as

$$n(\omega,\xi) = n_s - (n_s - n_\infty)(\cosh\omega - \cos\xi)^{1/2}\sum_{k=0}^{k=j-1}\left\{\frac{1}{\sqrt{\cosh\omega - \cos[\xi + 2k\pi j^{-1}]}} - \frac{1}{\sqrt{\cosh\omega - \cos[\xi + (2k-1)\pi j^{-1}]}}\right\} \quad (12)$$

Then, evaporation flux density is given by [1,4]

$$J = -D\frac{\cosh\omega - \cos\xi}{R}\frac{\partial n(\omega,\xi)}{\partial \xi} = \frac{(n_s - n_\infty)D}{R}(\cosh\omega - \cos\xi)\frac{\partial \psi(\omega,\xi)}{\partial \xi}, \quad (13)$$

where

$$\psi(\omega,\xi) = (\cosh\omega - \cos\xi)^{1/2} \sum_{k=0}^{k=j-1} \left\{ \frac{1}{\sqrt{\cosh\omega - \cos[\xi + 2k\pi j^{-1}]}} - \frac{1}{\sqrt{\cosh\omega - \cos[\xi + (2k-1)\pi j^{-1}]}} \right\}. \quad (14)$$

By differentiating in the formula (13), we get

$$\frac{\partial \psi(\omega,\xi)}{\partial \xi} = \frac{1}{2}(\cosh\omega - \cos\xi)^{-1/2} \sin\xi \sum_{k=0}^{k=j-1} \left\{ \frac{1}{\sqrt{\cosh\omega - \cos[\xi + 2k\pi j^{-1}]}} - \frac{1}{\sqrt{\cosh\omega - \cos[\xi + (2k-1)\pi j^{-1}]}} \right\} +$$

$$-\frac{1}{2}(\cosh\omega - \cos\xi)^{1/2} \sum_{k=0}^{k=j-1} \left\{ \frac{\sin[\xi + 2k\pi j^{-1}]}{(\cosh\omega - \cos[\xi + 2k\pi j^{-1}])^{\frac{3}{2}}} - \frac{\sin[\xi + (2k-1)\pi j^{-1}]}{(\cosh\omega - \cos[\xi + (2k-1)\pi j^{-1}])^{\frac{3}{2}}} \right\}$$

$$\sigma = \frac{V}{2R}(\cosh\omega - \cos\xi)^{1/2} \sin(0.5\pi j^{-1}) \sum_{k=0}^{k=j-1} \left\{ \frac{1}{\sqrt{\cosh\omega - \cos[\xi + 2k\pi j^{-1}]}} - \frac{1}{\sqrt{\cosh\omega - \cos[\xi + (2k-1)\pi j^{-1}]}} \right\} -$$

$$-\frac{V}{2R}(\cosh\omega - \cos\xi)^{3/2} \sum_{k=0}^{k=j-1} \left\{ \frac{\sin[\xi + 2k\pi j^{-1}]}{(\cosh\omega - \cos[\xi + 2k\pi j^{-1}])^{\frac{3}{2}}} - \frac{\sin[\xi + (2k-1)\pi j^{-1}]}{(\cosh\omega - \cos[\xi + (2k-1)\pi j^{-1}])^{\frac{3}{2}}} \right\}$$

Then

$$J = \frac{(n_s - n_\infty)D}{R}\tilde{J}, \quad (15)$$

where

$$\tilde{J} = (\cosh\omega - \cos\xi)^{1/2} \frac{\sin(0.5\pi j^{-1})}{2} \sum_{k=0}^{k=j-1} \left\{ \frac{1}{\sqrt{\cosh\omega - \cos[\xi + 2k\pi j^{-1}]}} - \frac{1}{\sqrt{\cosh\omega - \cos[\xi + (2k-1)\pi j^{-1}]}} \right\} -$$

$$-\frac{(\cosh\omega - \cos\xi)^{3/2}}{2} \sum_{k=0}^{k=j-1} \left\{ \frac{\sin[\xi + 2k\pi j^{-1}]}{(\cosh\omega - \cos[\xi + 2k\pi j^{-1}])^{\frac{3}{2}}} - \frac{\sin[\xi + (2k-1)\pi j^{-1}]}{(\cosh\omega - \cos[\xi + (2k-1)\pi j^{-1}])^{\frac{3}{2}}} \right\} \quad (16)$$

It can be shown that the first sum in formula (16) gives identically zero (can be verified by direct calculation), and the terms in brackets in the second sum are equal in absolute value. With this in mind, the expression is greatly simplified

$$\tilde{J} = (\cosh\omega - \cos\xi)^{3/2} \sum_{k=0}^{k=j-1} \frac{\sin[\xi + 2k\pi j^{-1}]}{(\cosh\omega - \cos[\xi + 2k\pi j^{-1}])^{\frac{3}{2}}}. \quad (17)$$

Taking into account (11), we get

$$\tilde{J}(j) = (\cosh\omega - \cos 0.5\pi j^{-1})^{3/2} \sum_{k=0}^{k=j-1} \frac{\sin[0.5\pi j^{-1} + 2k\pi j^{-1}]}{(\cosh\omega - \cos[0.5\pi j^{-1} + 2k\pi j^{-1}])^{\frac{3}{2}}}. \quad (18)$$

Thus, for any $j$, the evaporation flux density is given by

$$J(j) = \frac{(n_s - n_\infty)D}{R}(\cosh\omega - \cos 0.5\pi j^{-1})^{3/2} \sum_{k=0}^{k=j-1} \frac{\sin\left[0.5\pi j^{-1} + 2k\pi j^{-1}\right]}{\left(\cosh\omega - \cos\left[0.5\pi j^{-1} + 2k\pi j^{-1}\right]\right)^{\frac{3}{2}}}. \quad (19)$$

If k=j, the term under summation in (19) has the form

$$\frac{\sin\left[0.5\pi j^{-1} + 2j\pi j^{-1}\right]}{\left(\cosh\omega - \cos\left[0.5\pi j^{-1} + 2j\pi j^{-1}\right]\right)^{\frac{3}{2}}} = \frac{\sin\left[0.5\pi j^{-1}\right]}{\left(\cosh\omega - \cos\left[0.5\pi j^{-1}\right]\right)^{\frac{3}{2}}}. \quad (20)$$

If k=0, the term under summation in (19) has the same form

$$\frac{\sin\left[0.5\pi j^{-1}\right]}{\left(\cosh\omega - \cos\left[0.5\pi j^{-1}\right]\right)^{\frac{3}{2}}} = \frac{\sin\left[0.5\pi j^{-1}\right]}{\left(\cosh\omega - \cos\left[0.5\pi j^{-1}\right]\right)^{\frac{3}{2}}}. \quad (21)$$

Taking into account (20) and (21), equation (19) can be transformed as

$$J(j) = \frac{(n_s - n_\infty)D}{R}(\cosh\omega - \cos 0.5\pi j^{-1})^{3/2} \sum_{k=1}^{k=j} \frac{\sin\left[0.5\pi j^{-1} + 2k\pi j^{-1}\right]}{\left(\cosh\omega - \cos\left[0.5\pi j^{-1} + 2k\pi j^{-1}\right]\right)^{\frac{3}{2}}}. \quad (22)$$

Expression (22) is the equivalent to (19).

To apply expressions (19) or (22) for calculations, it is necessary to take into account the formula (11) and following geometric relationship [3-4]:

$$\cosh\omega = \frac{\sin^2\xi}{\cos\varphi + \cos\xi} + \cos\xi. \quad (23)$$

To establish a relationship between the parameter *j* and the corresponding contact angle θ, one has to use the geometric relation

$$\xi = \pi - \theta. \quad (24)$$

It means that

$$\theta = \pi\left(1 - \frac{1}{2j}\right) = \pi\frac{2j-1}{2j}, \text{ where } j=1,2,3\ldots \quad (25)$$

Table 1 shows the first two solutions corresponding to *j*=1 и *j*=2. It's obvious that

$$\lim_{j\to\infty}\theta = \pi. \quad (26)$$

First three solution of the expression (22) (*j*=1,2,3) placed into Table 1.

## 3. Examples of exact solutions

Using Table 1, one can compare the solutions for contact angles 135° and 150° obtained from (19) with the general solution $f(\theta)$ that works for arbitrary contact angles given by equation (5):

$$f_{\theta=135°}(\varphi) = \frac{1}{\sqrt{2}} - \frac{0.25}{\left(\frac{3}{2} + \sqrt{2}\cos\varphi\right)^{\frac{3}{2}}}, \quad (27)$$

$$f_{\theta=150°}(\varphi) = \frac{1}{2}\left(1 + \left(7 + 4\sqrt{3}\cos\varphi\right)^{-\frac{3}{2}} - 2\left(4 + 2\sqrt{3}\cos\varphi\right)^{-\frac{3}{2}}\right). \tag{28}$$

**Table 1**.

*First three solutions for evaporation flux density J (contact angles θ of 90º, 135º and 150º)*

| j | θ | J |
|---|---|---|
| 1 | $\pi\left(1-\frac{1}{2}\right) = \frac{\pi}{2} = 90°$ | $J = \dfrac{(n_s - n_\infty)D}{R}$ |
| 2 | $\pi\left(1-\frac{1}{4}\right) = \frac{3\pi}{4} = 135°$ | $J = \dfrac{(n_s - n_\infty)D}{R}\left(\dfrac{1}{\sqrt{2}} - \dfrac{0.25}{\left(\frac{3}{2} + \sqrt{2}\cos\varphi\right)^{\frac{3}{2}}}\right)$ |
| 3 | $\pi\left(1-\frac{1}{6}\right) = \frac{5\pi}{6} = 150°$ | $J = \dfrac{(n_s - n_\infty)D}{2R}\left(1 + \left(7 + 4\sqrt{3}\cos\varphi\right)^{-\frac{3}{2}} - 2\left(4 + 2\sqrt{3}\cos\varphi\right)^{-\frac{3}{2}}\right)$ |

Figure 2 represents the dependences of the evaporation flux density (dimensionless) on the polar angle, which is given by formulas (27) and (28).

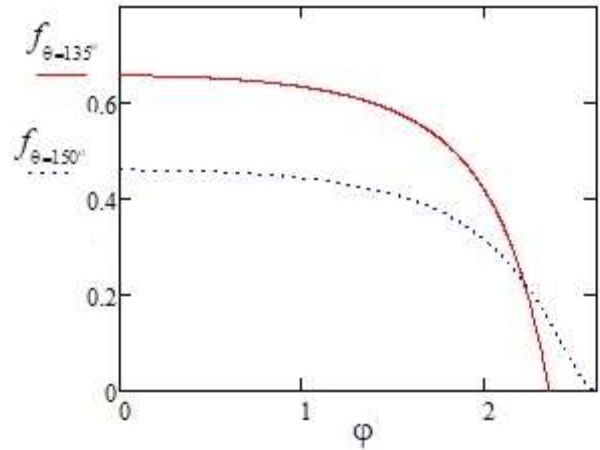

**Figure 2.** Graphs of the functions represented by formulas (27) and (28).

It is easy to verify by direct calculation that formula (5) gives the same curves, which confirms the correctness of the mathematical transformation that led to the formula (19) or the expression (22).

## 4. Conclusion

New complex solution (22) was derived for the evaporation flux density of a small liquid droplet having the shape of an axisymmetric spherical segment deposited on a horizontal substrate for the set of discrete contact angles $\theta = \pi\left(1 - \frac{1}{2j}\right)$, where $j$=1,2,3… As an example, very simple exact expressions (27) and (28) were obtained explicitly for the evaporation flux density for droplets with contact angles $\frac{3\pi}{4} = 135°$ and $\frac{5\pi}{6} = 150°$ that do not contain integral dependencies. They can also be used as approximate expressions for a narrow range of contact angles around the specified values.

The work was supported by the Federal Scientific Research Center "Crystallography and Photonics" RAS.